 \documentclass[smallabstract,smallcaptions]{dccpaper}

\usepackage{epsfig}
\usepackage{amsmath}
\usepackage{amssymb}
\usepackage{color}
\usepackage{url}
\usepackage{tikz}
\usetikzlibrary{arrows.meta, automata,
                calc,
                positioning,
                quotes}
\usepackage{placeins}
\usepackage{pgfplots}
\usepackage{multicol}
\usepackage{multirow}
\usepackage{array,graphicx}
\usepackage{mathtools,booktabs}
\usepackage{array}
\usepackage{siunitx}
\usepackage{xcolor}

\usepackage{enumitem}
\definecolor{darkbluishgray}{RGB}{40,40,60}
\definecolor{bluish}{RGB}{80,80,120}
\definecolor{blue_corr_1}{RGB}{10,100,200}
\definecolor{pink_heatmap}{RGB}{196,112,112}
\definecolor{pink_heatmap_m}{RGB}{150,92,92}
\definecolor{yellow_heatmap}{RGB}{240,235,179}
\definecolor{yellow_heatmap_m}{RGB}{200,195,139}
\newcommand{\vxv}[2]{$#1\!\times\!#2$}

\newlength{\figurewidth}
\newlength{\smallfigurewidth}

\begin{document}

\title
{\large
\textbf{Entropy Coding for Non-Rectangular Transform Blocks using Partitioned DCT Dictionaries for AV1}
}

\author{%
Priyanka Das, Tim Classen, and Mathias Wien\\[0.5em]
{\small\begin{minipage}{\linewidth}\begin{center}
\begin{tabular}{ccc}
$^{\ast}$Lehrstuhl für Bildgebung und Bildverarbeitung &&\\
RWTH Aachen University && \\
\url{{priyanka.das,tim.classen,mathias.wien}@lfb.rwth-aachen.de} &&
\end{tabular}
\end{center}\end{minipage}}
}

\maketitle
\thispagestyle{empty}

\begin{abstract}
Recent video codecs such as VVC and AV1 apply a  Non-rectangular (NR) partitioning to combine prediction signals using a smooth blending around the boundary, followed by a rectangular transform on the whole block. The NR signal transformation is not yet supported. A transformation technique that applies the same partitioning to the 2D Discrete Cosine Transform (DCT) bases and finds a sparse representation of the NR signal in such a dictionary showed promising gains in an experimental setup outside the reference software. This method uses the regular inverse transformation at the decoder to reconstruct a rectangular signal and discards the signal outside the region of interest. This design is appealing due to the minimal changes required at the decoder. However, current entropy coding schemes are not well-suited for optimally encoding these coefficients because they are primarily designed for DCT coefficients. This work introduces an entropy coding method that efficiently codes these transform coefficients by effectively modeling their properties. The design offers significant theoretical rate savings, estimated using conditional entropy, particularly for scenarios that are more dissimilar to DCT in an experimental setup.
\end{abstract}

\Section{Introduction}
Traditionally, video codecs apply prediction, transformation (TX) and entropy coding on a rectangular block basis. A sinusoidal TX is applied to the full or subblock of the error signal after block prediction. Recently, AOMedia Video 1 (AV1) \cite{9363937} codec introduced a compound tool known as Wedge mode, which combines two inter-inter or inter-intra prediction signals. This mode assigns the prediction signals to two NR regions by dividing a block using oblique angles. This mode uses an 8-pixel wide blending area around the prediction boundary. Similarly, Geometric Partition Mode (GPM) in VVC combines two inter-prediction signals using an adaptive blending. Both modes apply rectangular TX to the combined NR residuals. \par 

Currently, these modes provide an approximate NR partitioning for prediction signals. Enabling a true NR partitioning involves several additional steps: 1) applying a sharper transition at the NR prediction boundary, 2) performing TX on both NR blocks separately, 3) optimizing entropy coding for NR TX, and 4) adapting the deblocking filter to be usable at the NR boundary. Furthermore, AV1 includes a model-based rate estimation algorithm to reduce encoder complexity, which is trained on DCT coefficients. This algorithm should also be updated to accommodate the NR TX as well.
Among available arbitrarily shaped TXs \cite{350781,678616,Gilge1989,6975156,7126941,744271,413280,EI1998,344933}, two techniques \cite{344933} and \cite{678616} have received recent interest as an NR TX \cite{9191301}, \cite{8456238}. The scheme described in \cite{9191301} demonstrated promising gains in an experimental setup. The technique utilizes the existing inverse rectangular TX block at the decoder to reconstruct a rectangular signal and discard the signal outside the NR region of interest. This design is particularly desirable due to the minimal changes required at the decoder.
However, the corresponding TX coefficient distribution differs from the distribution which is observed with DCT-transform. This TX technique uses a sparse representation method along with an overcomplete dictionary, where the dictionary elements, also known as atoms are correlated with their neighbors. This results in non-zero TX coefficients surrounded by zeros. Thereby, although the TX coefficients have high energy concentration in the top-left corner, similar to DCT, their spatial compaction is significantly lower. Furthermore, similar to DCT, a weak correlation is observed between non-zero TX coefficients within a neighborhood. The existing entropy coding methods \cite{8296882},\cite{1218195} in the codecs are tailored to DCT properties and are not optimally suited for them. \par

This work investigates an entropy coding scheme to enable true NR partitioning in AV1, using the NR TX from \cite{9191301} with minimal design changes. The scheme efficiently encodes sparse TX coefficients, where a weak correlation is observed in neighboring non-zero TX coefficients and a strong correlation exists among neighboring atoms. The final algorithm presented here is tailored to AV1 and NR-TX described in \cite{9191301}. However, the general principles of the presented entropy coding design can be used to encode sparse TX coefficients based on the properties of the corresponding dictionary.\par

\Section{Related Work}\label{Sec:Relatedwork}
\SubSection{Entropy Coding}
Entropy measures the theoretical lower bound of the achievable rate for a given set of source symbols. Most popular codecs apply prediction, TX, and quantization to achieve a low-entropy block of integer coefficients to be transmitted. These blocks are encoded into the bitstream using entropy coding schemes such as run-length, Variable length (VLC), Huffman, or Arithmetic coding. The methods presented in \cite{tudor1995mpeg,ITU-H263-2005,ISO-MPEG4} use a combination of run-length and VLC. However, VLC uses fixed, pre-defined tables that do not adapt to the actual symbol statistics. The fixed connection between codeword and symbol has been relaxed later in Context Adaptive VLCs. \par
\begin{figure}[t]
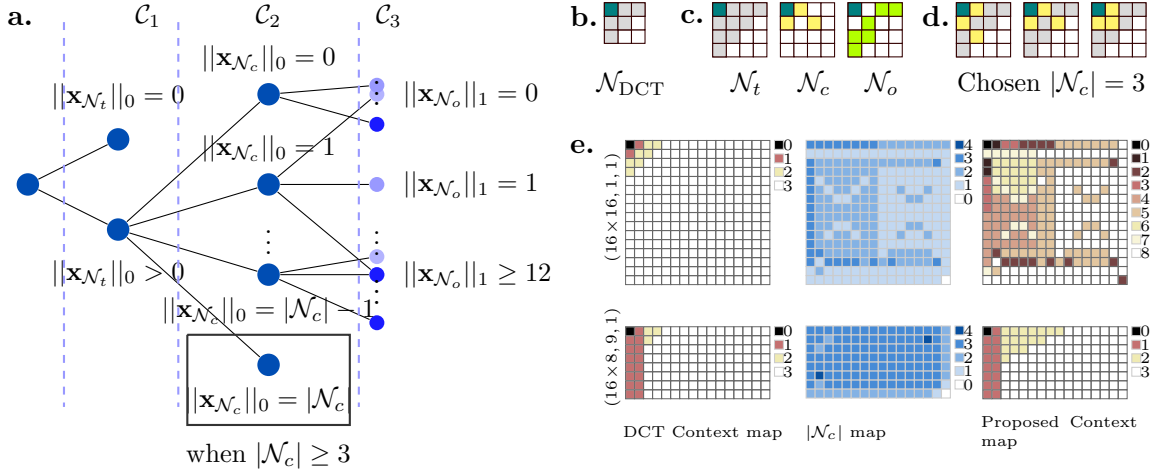

 \begin{center}
 \begin{tabular}{p{.5cm}p{6.5cm}p{6.5cm}}
  \hspace*{-.2cm}\raisebox{1.1cm}{\small{\textbf{a.}}}&
\vspace*{-1.5cm}
\hspace*{-1cm}\multirow{2}{*}{\begin{tikzpicture}[scale=.8]
[level distance=20mm,
level 1/.style={sibling distance=10mm},
level 2/.style={sibling distance=15mm},
level 3/.style={sibling distance=7mm}]
\tikzset{treenode/.style={fill=blue!70!green,circle,inner sep=3pt}}
\tikzset{leaf/.style={fill=blue!70!green,circle,inner sep=2pt}}
\tikzset{leaf1/.style={fill=blue!30!white,circle,inner sep=2pt}}
\tikzset{leaf2/.style={fill=blue!40!white,circle,inner sep=2pt}}
\tikzset{leaf3/.style={fill=blue!90!white,circle,inner sep=2pt}}

    \node [treenode] at (0,0){} [grow=right] 
    child[level distance=15mm] {node[treenode](b) {}
      child[level distance=25mm,sibling distance=15mm] {node[treenode](e) {}}
      child[level distance=25mm,sibling distance=15mm] {node[treenode](f) {}
              child[level distance=18mm,sibling distance=8mm] {node[leaf3](g3) {}}
              child[level distance=18mm,sibling distance=3mm] {node[leaf2](i3) {}}
              child[level distance=18mm,sibling distance=3mm] {node[leaf1](j3) {}}}
      child[level distance=25mm,sibling distance=15mm] {node[treenode](d) {}
              child[level distance=18mm] {node[leaf3](g2) {}}
              child[level distance=18mm] {node[leaf2](i2) {}}
              child[level distance=18mm] {node[leaf1](j2) {}}}
      child[level distance=25mm,sibling distance=15mm] {node[treenode](c) {} 
              child[level distance=18mm,sibling distance=10mm] {node[leaf3](g1) {}}
              child[level distance=18mm,sibling distance=3mm] {node[leaf2](i1) {}}
      }
    }
    child[level distance=15mm] {node[treenode](a) {}};
    \node[above = 1mm of a] {\footnotesize{$||\mathbf{x}_{\mathcal{N}_t}||_0=0$}};
    \node[below = 1mm of b] {\footnotesize{$||\mathbf{x}_{\mathcal{N}_t}||_0>0$}};
    

    \node[above = .05mm of c] {\footnotesize{$||\mathbf{x}_{\mathcal{N}_c}||_0=0$}};
    \node[above = .05mm of d] {\footnotesize{$||\mathbf{x}_{\mathcal{N}_c}||_0=1$}};
    \node[below = 1mm of d] {\footnotesize{$\vdots$}};
    \node[below = .01mm of e] {\footnotesize{$||\mathbf{x}_{\mathcal{N}_c}||_0=|\mathcal{N}_c|$}};
    \node[below = .05mm of f] {\footnotesize{$||\mathbf{x}_{\mathcal{N}_c}||_0=|\mathcal{N}_c|-1$}};


    \node[above = .01mm of g1] {$\vdots$};
    \node[above = .01mm of g3] {$\vdots$};

    \node[right = 1mm of j2] {\footnotesize{$||\mathbf{x}_{\mathcal{N}_o}||_1=0$}};
    \node[right = 1mm of i2] {\footnotesize{$||\mathbf{x}_{\mathcal{N}_o}||_1=1$}};
    \node[above = .1mm of g2]{\footnotesize{$\vdots$}};
    \node[right = 1mm of g2] {\footnotesize{$||\mathbf{x}_{\mathcal{N}_o}||_1\geq12$}};

    \draw[blue!40!white,thick,dashed] (.6,2.7) -- (.6,-3.7);
    \draw[blue!40!white,thick,dashed] (2.5,2.7) -- (2.5,-3.7);
    \draw[blue!40!white,thick,dashed] (5.5,2.7) -- (5.5,-3.7);

    \node at (2,2.8) {\footnotesize{\textbf{$\mathcal{C}_1$}}};
    \node at (4,2.8) {\footnotesize{\textbf{$\mathcal{C}_2$}}};
    \node at (6,2.8) {\footnotesize{\textbf{$\mathcal{C}_3$}}};

    \draw[white!20!black,thick] (2.65,-2.5) -- (5.35,-2.5) --  (5.35,-4) -- (2.65,-4) -- (2.65,-2.5);
    \node at (4,-4.5) {\footnotesize{when $|\mathcal{N}_c|\geq3$}};
\end{tikzpicture}} & \hspace*{-.7cm}\begin{tikzpicture}[scale=.18]
\node at (-1.5,3) {\small{\textbf{b.}}};
       \foreach \y in {0,1,2}
	{
              \foreach \x in {0,1,2}
              {
		\filldraw[fill=white, draw=red!20!black](\x,\y+1) rectangle (\x+1,\y+2); 
              }
	} 
       \foreach \y in {0,1,2}
	{
              \foreach \x in {0,...,\y}
              {
		\filldraw[fill=black!15!white, draw=red!20!black](\x,\y+1) rectangle (\x+1,\y+2); 
              }
	} 
	\filldraw[fill=blue!50!green, draw=red!20!black](0,3) rectangle (1,4); 
	\node at (2,-2) (a) {\footnotesize{$\mathcal{N}_{\mathrm{DCT}}$}};

\node at (6.5,3) {\small{\textbf{c.}}};
       \foreach \y in {0,1,2,3}
	{
              \foreach \x in {0,1,2,3}
              {
		\filldraw[fill=white, draw=red!20!black](\x+8,\y) rectangle (\x+9,\y+1); 
              }
	} 
       \foreach \y in {0,1,2,3}
	{
              \foreach \x in {0,...,\y}
              {
		\filldraw[fill=black!15!white, draw=red!20!black](\x+8,\y) rectangle (\x+9,\y+1); 
              }
	} 
	\filldraw[fill=blue!50!green, draw=red!20!black](8,4) rectangle (9,3); 
	\node at (10.5,-2) (a) {\footnotesize{$\mathcal{N}_t$}};


		\foreach \y in {0,1,2,3}
		{
                     \foreach \x in {0,1,2,3}
                     {
			\filldraw[fill=white, draw=red!20!black](5+8+\x,\y) rectangle (\x+8+6,\y+1); 
                     }
		} 
		\filldraw[fill=blue!50!green, draw=red!20!black](5+8,4) rectangle (6+8,3); 
		\filldraw[fill=yellow!70!white, draw=red!20!black](5+8,3) rectangle (6+8,2); 
		\filldraw[fill=yellow!70!white, draw=red!20!black](6+8,4) rectangle (7+8,3); 
              \filldraw[fill=yellow!70!white, draw=red!20!black](7+8,3) rectangle (8+8,2);
		\node at (7.5+8,-2) (a) {\footnotesize{$\mathcal{N}_c$}};


		\foreach \y in {0,1,2,3}
		{
                     \foreach \x in {0,1,2,3}
                     {
			\filldraw[fill=white, draw=red!20!black](10+8+\x,\y) rectangle (\x+8+11,\y+1); 
                     }
		} 
              \foreach \y in {0,1,2,3}
		{
                     \foreach \x in {0,...,\y}
                     {
			\filldraw[fill=green!30!yellow, draw=red!20!black](10+8+\x,\y) rectangle (\x+8+11,\y+1); 
                     }
		} 
		\filldraw[fill=blue!50!green, draw=red!20!black](10+8,4) rectangle (11+8,3); 
		\filldraw[fill=white, draw=red!20!black](10+8,3) rectangle (11+8,2); 
		\filldraw[fill=white, draw=red!20!black](11+8,4) rectangle (12+8,3); 
              \filldraw[fill=white, draw=red!20!black](12+8,3) rectangle (13+8,2);
		\node at (12.5+8,-2) (a) {\footnotesize{$\mathcal{N}_o$}};
		\vspace*{.3cm}

\node at (24.5,3) {\small{\textbf{d.}}};

              \foreach \y in {0,1,2,3}
		{
                     \foreach \x in {0,1,2,3}
                     {
			\filldraw[fill=white, draw=red!20!black](\x+26,\y) rectangle (\x+1+26,\y+1); 
                     }
		} 
              \foreach \y in {0,1,2,3}
		{
                     \foreach \x in {0,...,\y}
                     {
			\filldraw[fill=black!15!white, draw=red!20!black](\x+26,\y) rectangle (\x+1+26,\y+1); 
                     }
		} 
		\filldraw[fill=blue!50!green, draw=red!20!black](0+26,4) rectangle (1+26,3); 
		\filldraw[fill=yellow!70!white, draw=red!20!black](0+26,3) rectangle (1+26,2); 
		\filldraw[fill=yellow!70!white, draw=red!20!black](1+26,4) rectangle (2+26,3); 
              \filldraw[fill=yellow!70!white, draw=red!20!black](1+26,2) rectangle (2+26,1);

		\foreach \y in {0,1,2,3}
		{
                     \foreach \x in {0,1,2,3}
                     {
			\filldraw[fill=white, draw=red!20!black](5+\x+26,\y) rectangle (\x+6+26,\y+1); 
                     }
		} 
              \foreach \y in {0,1,2,3}
		{
                     \foreach \x in {0,...,\y}
                     {
			\filldraw[fill=black!15!white, draw=red!20!black](5+\x+26,\y) rectangle (\x+6+26,\y+1); 
                     }
		} 
		\filldraw[fill=blue!50!green, draw=red!20!black](5+26,4) rectangle (6+26,3); 
		\filldraw[fill=yellow!70!white, draw=red!20!black](5+26,3) rectangle (6+26,2); 
		\filldraw[fill=yellow!70!white, draw=red!20!black](6+26,4) rectangle (7+26,3); 
              \filldraw[fill=yellow!70!white, draw=red!20!black](7+26,3) rectangle (8+26,2);

		\foreach \y in {0,1,2,3}
		{
                     \foreach \x in {0,1,2,3}
                     {
			\filldraw[fill=white, draw=red!20!black](10+\x+26,\y) rectangle (\x+11+26,\y+1); 
                     }
		} 
              \foreach \y in {0,1,2,3}
		{
                     \foreach \x in {0,...,\y}
                     {
			\filldraw[fill=black!15!white, draw=red!20!black](10+\x+26,\y) rectangle (\x+11+26,\y+1); 
                     }
		} 
		\filldraw[fill=blue!50!green, draw=red!20!black](10+26,4) rectangle (11+26,3); 
		\filldraw[fill=yellow!70!white, draw=red!20!black](10+26,3) rectangle (11+26,2); 
		\filldraw[fill=yellow!70!white, draw=red!20!black](11+26,4) rectangle (12+26,3); 
              \filldraw[fill=yellow!70!white, draw=red!20!black](11+26,3) rectangle (12+26,2);
		\node at (6+27,-2) (a) {\footnotesize{Chosen $|\mathcal{N}_c| = 3$}};
		\vspace*{.3cm}

\end{tikzpicture}\\
&& \hspace*{-.7cm}\begin{tikzpicture}
        \node at (-.5,2.3) {\small{\textbf{e.}}};

        \node at (-.1, 1.5) {\rotatebox[origin=c]{90}{\tiny{(\vxv{16}{16}$,1,1$)}}};
        \node at (1.25, 1.5) {\input{figures/DCT_ctxmap.tikz}};
        \node at (3.65, 1.5) {\input{figures/16x16x1_corrmap.tikz}};
        \node at (6, 1.5) {\input{figures/proposed_16x16x1.tikz}};

        \node at ( -.1, -.5) {\rotatebox[origin=c]{90}{\tiny{(\vxv{16}{8}$,9,1$)}}};
        \node at (1.25, -.5) {\input{figures/DCT_16x8_ctx_map.tikz}};
        \node at (3.65, -.5) {\input{figures/16x8x9_corrmap.tikz}};
        \node at (6, -.5) {\input{figures/proposed_16x8.tikz}};

        \node[align=left] at (1.1, -1.5) {\tiny{DCT Context map}};
        \node[align=left] at (3, -1.5) {\tiny{$|\mathcal{N}_c|$ map}};
        \node[align=left,text width=3cm] at (6.3, -1.5) {\parbox{2.2cm}{\tiny{Proposed Context map}}};

\end{tikzpicture}\\
 \end{tabular}
 \end{center}
 \caption{\label{fig:example}%
 \small{a. Full context tree for entropy coding of non-rectangular transform coefficients, b. $\mathcal{N}_t$ for DCT, c. Definition of $\mathcal{N}$ types, d. For CT-s, e. Context map used for DCT, $|\mathcal{N}_c|$ map, Designed context map }}
 \end{figure}

In contrast, arithmetic coders do not require a pre-defined table. They use decision rules and a well-defined state at the arrival of a new symbol, allowing the decoder to reconstruct a group of symbols whenever possible. Arithmetic coders were initially designed for codecs in \cite{gonzales1989dct,ITU-H263-2005,marpe1999very,heising2001wavelet,choi1999motion}. An arithmetic coder easily accommodates inter-symbol redundancies by defining contexts and using conditional probabilities to approach conditional entropy. All modern codecs employ such arithmetic coding with symbol adaptivity, such as CABAC \cite{1218195} and level-map $M$-ary arithmetic coding in AV1 \cite{8296882}. Since this work is designed based on AV1, the corresponding entropy coding scheme is summarized in the following. 
\SubSection{Entropy Coding scheme in AV1}
\label{sec:av1entropycoding}
AV1 \cite{9363937} employs an adaptive $M$-ary symbol arithmetic coding scheme with a level-map approach to encode quantized TX coefficients block. The block is first mapped into a 1D vector using a scan. Three different scans (zig-zag, column and row) are employed based on TX (2D, vertical 1D, horizontal 1D). They are decomposed into 4 symbols: sign bit, base range (BR) \{$0,1,2,>2$\}, low range (LR) \{$0,1,2,>2$\}, high range $[0,2^{15}]$. These symbols are encoded using an $M$-ary arithmetic coding and an adaptive probability model. The contexts for the probability models are designed separately for BR and LR symbols by exploiting the high correlation of DCT coefficients with their spatial neighbors and coefficients in the same frequency band. A neighborhood of 5 (illustrated in Fig. \ref{fig:example}-b below) and 3 coefficients are considered for BR and LR symbols respectively for 2D TX. The sum of the absolute value of the coefficients in this neighborhood is categorized into ranges $[0,\{1,2\},\{3,4\},\{5,6\},\{>=7\}]$ to establish contexts for the BR symbol AC coefficients. To leverage the weak correlation among coefficients within the same frequency bands, the block is divided into regions where the same contexts are used. Fig. \ref{fig:example}-e below illustrates examples of two block sizes. Contexts are also shared across different block sizes. In total, there are 22 different contexts, including one for the DC coefficient. The updating rate of the probability model depends on how frequently a symbol occurs at the frame level.
\SubSection{Transformation technique:}
The TX technique described in \cite{9191301} utilizes a partitioned DCT dictionary to transform an NR signal. The corresponding NR region from the nearest larger rectangular DCT serves as the basis for constructing the dictionary. The number of atoms in such dictionary equals the bounding box area. Next, the Orthogonal Matching Pursuit (OMP) \cite{pati1993orthogonal} is used to obtain a sparse representation of the NR signal within the overcomplete dictionary. The method presented in \cite{9191301} uses this representation to linearly combine the unused part of the DCT  to extend the shape of the NR signal to a rectangular one. A regular DCT is then applied on the entire block to derive the TX coefficients.\par
The key advantage of this method is that the inverse TX step at the decoder remains almost untouched. A regular inverse DCT is applied on the TX coefficient block, and only the the region of interest is considered for reconstruction. If the sparse representation has negligible loss, it can be directly used as TX coefficients with appropriate scaling eliminating the need for further extension mechanism. In contrast, for lossy OMP approximations, the extension technique will provide better quality. While a near-lossless OMP approximation, the extension technique provided slight gain in high bitrates, it also added more complexity. Due to this, the scaling method is employed in the current work. 
\begin{figure}[tb]
  \centering
\setlength{\abovedisplayskip}{0pt}
\setlength{\belowdisplayskip}{0pt}
\setlength{\abovedisplayshortskip}{0pt}
\setlength{\belowdisplayshortskip}{0pt}
   \setlength{\tabcolsep}{5pt}
 	\renewcommand{\arraystretch}{1}
 	\begin{tabular}{p{1.5cm}|p{1.5cm}|p{1.5cm}p{1.5cm}|p{1.5cm}|p{1.5cm}p{.2cm}}
   \raisebox{-.2cm}{     \begin{tikzpicture}[scale=0.22]
   \newcommand\Square[1]{+(-#1,-#1) rectangle +(#1,#1)}

  \foreach \y [count=\n] in { 
    {35,     9,    17,    13,    28},
    { 9,    36,    64,    18,    12},
    {17,    64,   100,    52,    11},
    {13,    18,    52,    20,     5},
    {28,    12,    11,     6,    13},
      } {
      \foreach \x [count=\m] in \y {
         \draw[gray!30!,fill=black!\x!white] (\m,-\n) \Square{14pt} ; 

      }
    }
    \draw[black] (.5,0) rectangle (4.5,2);
    \draw[fill=darkbluishgray] (.5,.5) -- (3.5,2) -- (.5,2) -- (.5,.5);
    \node at (3,-6.7) {\footnotesize{Type 1}};
\end{tikzpicture} } &\raisebox{-.2cm}{\begin{tikzpicture}[scale=0.22]
   \newcommand\Square[1]{+(-#1,-#1) rectangle +(#1,#1)}

  \foreach \y [count=\n] in { 
    { 0,     7,     0,    15,     0},
    { 9,     0,    35,     0,     9},
    { 0,    38,   100,    45,     0},
    {17,     0,    42,     0,    18},
    { 0,     6,     0,    14,     0},
    } {
      \foreach \x [count=\m] in \y {
         \draw[gray!30!,fill=black!\x!white] (\m,-\n) \Square{14pt} ; 

      }
    }
    \draw[black] (.5,0) rectangle (4.5,2);
    \draw[fill=darkbluishgray] (.5,0) -- (4.5,2) -- (.5,2) -- (.5,0);
    \node at (3,-6.7) {\footnotesize{Type 2}};
\end{tikzpicture} } &\raisebox{-.2cm}{\begin{tikzpicture}[scale=0.22]
   \newcommand\Square[1]{+(-#1,-#1) rectangle +(#1,#1)}

  \foreach \y [count=\n] in { 
   { 0,    23,     0,    22,     0},
   { 7,     0,    51,     0,     4},
   { 0,    25,   100,    26,     0},
   {14,     0,    56,     0,    10},
   { 0,    10,     0,    15,     0},
} {
      \foreach \x [count=\m] in \y {
         \draw[gray!30!,fill=black!\x!white] (\m,-\n) \Square{14pt} ; 

      }
    }
    \draw[black] (.5,0) rectangle (4.5,4);
    \draw[fill=darkbluishgray] (.5,1) -- (4.5,3) --(4.5,4) -- (.5,4) -- (.5,1);
    \node at (3,-6.7) {\footnotesize{Type 3}};

\end{tikzpicture}} &\raisebox{-.2cm}{   \begin{tikzpicture}[scale=0.22]
   \newcommand\Square[1]{+(-#1,-#1) rectangle +(#1,#1)}

  \foreach \y [count=\n] in { 
     { 0,     7,     0,    14,     0},
    {23,     0,    25,     0,    10},
    { 0,    51,   100,    56,     0},
    {22,     0,    26,     0,    15},
    { 0,     4,     0,    10,     0},
    } {
      \foreach \x [count=\m] in \y {
         \draw[gray!30!,fill=black!\x!white] (\m,-\n) \Square{14pt} ; 
      }
    }
    \draw[black] (.5,0) rectangle (4.5,4);
    \draw[fill=darkbluishgray] (.5,0) -- (1.5,0) -- (3.5,4) --(.5,4) -- (.5,0);

    \node at (3,-6.7) {\footnotesize{Type 3}};
\end{tikzpicture}} &\raisebox{-.2cm}{         \begin{tikzpicture}[scale=0.22]
   \newcommand\Square[1]{+(-#1,-#1) rectangle +(#1,#1)}

  \foreach \y [count=\n] in { 
    { 9,    12,     6,     5,    12},
    {12,     8,    31,     9,     5},
    { 6,    30,   100,    22,     4},
    { 5,     9,    23,     7,     4},
    {12,     5,     4,     4,     5},
     } {
      \foreach \x [count=\m] in \y {
         \draw[gray!30!,fill=black!\x!white] (\m,-\n) \Square{14pt} ; 

      }
    }
    \draw[black] (.5,0) rectangle (4.5,2);
    \draw[fill=darkbluishgray] (.5,0)--(1.5,0) -- (4.5,1.5) -- (4.5,2) -- (.5,2)--(.5,0);
    \node at (3,-6.7) {\footnotesize{Type 4}};
\end{tikzpicture} } &\raisebox{-.2cm}{    \begin{tikzpicture}[scale=0.22]
   \newcommand\Square[1]{+(-#1,-#1) rectangle +(#1,#1)}

  \foreach \y [count=\n] in { 
     {1,     4,    0,     1,     2},
     {0,     0,    2,     2,     0},
     {26,    40,   100,   26,     17},
     {0,     0,    2,     2,     0},
     {2,     3,    0,     2,     2},
     } {
      \foreach \x [count=\m] in \y {
         \draw[gray!30!,fill=black!\x!white] (\m,-\n) \Square{14pt} ; 

      }
    }
    \draw[black] (.5,0) rectangle (4.5,2);
   \draw[fill=darkbluishgray] (.5,0)--(3,0) -- (4,2) -- (.5,2) --(.5,0);
   \node at (3,-6.7) {\footnotesize{Type 5}};
\end{tikzpicture} }&\raisebox{.15cm}{\begin{tikzpicture}[xscale=.3,yscale=.22]
\draw[white,fill, top color=black, bottom color=white] (0,-4) -- (.4,-4) -- (.4,1) -- (0,1) -- (0,-4);
\draw[] (-.1,-4) --(.5,-4);
\node at (1.1,-4) {\tiny{0}};
\draw[] (-.1,1) --(.5,1);
\node at (1.1,1) {\tiny{1}};
\draw[] (-.1,-1.5) --(.5,-1.5);
\node at (1.1,-1.5) {\tiny{0.5}};
\end{tikzpicture} }\\
\noalign{\hrule height .5pt}
  $\tiny{\begin{aligned}[t]
  ( 8,16)&: 8\\
  (16,32)&: 8\\
  ( 8,32)&: 8
\end{aligned}}$
    &
    $\tiny{\begin{aligned}[t]
  (\; 4, \;8)&:\;8\\
  (\;8,16)&:24\\
  (16,32)&:16
\end{aligned}}$ &\multicolumn{2}{c|}{$\tiny{\begin{aligned}[t]
  (\; 8,\; 8)&:16\\
  (16,16)&:16\\
  (32,32)&:16\\
  (\; 8,16)&:24\\
  (16,32)&:\;8
\end{aligned}}$}  &  $\tiny{\begin{aligned}[t]
  (\; 8, \;8)&: 8\\
  (16,16)&:8\\
  (32,32)&:8\\
  ( \;8,16)&:8\\
  (16,32)&:8\\
  (\;8,32)&:8\\
\end{aligned}}$ &$\tiny{\begin{aligned}[t]
  (\; 8, 16)&: 8\\
  ( 16,32)&: 8\\
\end{aligned}}$&\\


\bottomrule
  \end{tabular}	
  \caption{Top: NR shapes for different types, example of correlation of an atom at position (3,3) in a \vxv{5}{5} neighborhood for each type. Below: Number of cases of each NR shape, characterized by blocksize and type} 
  \label{fig:nbd_corr}
\end{figure}
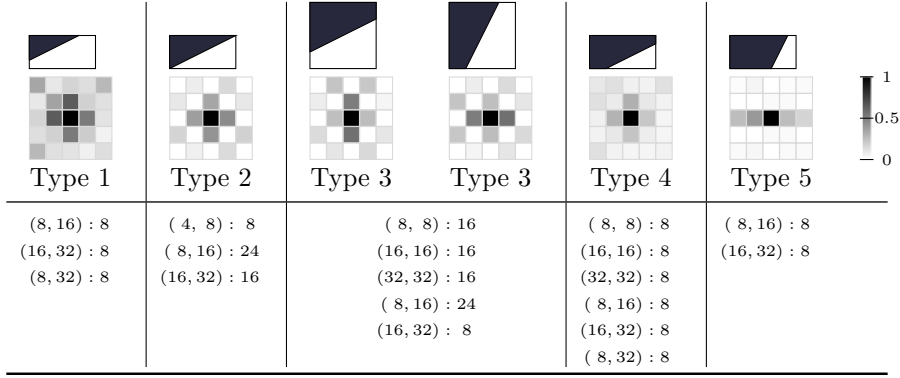 
\SubSection{NR shapes in AV1 wedge mode}
\label{sec:NRshape}
AV1 adopted 16 wedges using angles \{\ang{0}, \ang{27}, \ang{63}, \ang{90}, \ang{117}, \ang{153}\} for 9 block sizes with dimensions \{8, 16, 32\}. Each wedge mode generates 2 NR regions. Excluding the 72 rectangular shapes, there are a total of 216 NR regions; however, these are not all unique. A bounding box with dimensions $2^m, m\in N$ is applied around an NR region.  Some NR regions of aspect ratio 1:4 can be subdivided into one rectangular and one NR region. By employing rotation, mirroring, and transposition on these shapes, a set of 19 unique NR blocks can be derived.\par
Two properties of the NR shapes are; 1) $r_a:$ the ratio between the area of the NR region and the bounding box. $\text{2) SH:}$ presence of diagonal symmetry or closer to rectangles due to longer parallel lines. Based on these properties, the blocks can be categorized into the following 5 types. Each type is exemplified with parameters in the form of (\vxv{\text{width}}{\text{height}}, $\text{wedge index}$, $\text{region}$). This form is used in the rest of the paper to specify an NR block.
\begin{enumerate}[itemsep=1.5pt, label={Type \arabic*:}, leftmargin=1.6cm]
  \item $r_a\approx\frac{1}{4}$, (e.g. \vxv{16}{8}, 9, 1) 
  \item $r_a=\frac{1}{2}$, SH: diagonal symmetry (e.g. \vxv{8}{16}, 2, 1) 
  \item $r_a=\frac{1}{2}$, SH: closer to rectangle (e.g. \vxv{8}{16}, 1, 1)
  \item $r_a\approx\frac{3}{4}$, SH: diagonal symmetry (e.g. \vxv{8}{16}, 14, 1)
  \item $r_a\approx\frac{3}{4}$, SH: closer to rectangle(e.g. \vxv{8}{16}, 10, 1)
 \end{enumerate}
Fig. \ref{fig:nbd_corr} gives a complete list of the number of cases for these 19 unique NR shapes characterized by their block type and size. The blocksize ($w,h$) in the figure refers to both \vxv{w}{h} and \vxv{h}{w}. $92\%$ of the NR shapes are either square or 1:2. Only the blocks with 1:2 aspect ratio cover all the block types.





\SubSection{Dictionary Analysis:}
The TX coefficient properties are linked directly to the properties of the dictionary. The dictionaries are created by partitioning DCT bases, resulting in one dictionary for each unique NR shape. This section discusses the unique correlation pattern for each dictionary type, which is the key property linked to the TX coefficients' behaviour. The $r_a$ property of a block is inversely proportional to the dictionary's overcompleteness. Higher $r_a$ values and $sh$ closer to a rectangle generate dictionaries closer to the DCT. These overcomplete dictionaries have a correlation between neighboring atoms. The correlation is found to be negligible if the $l_1$ distance between atoms is higher than 5. The inter-atom correlation distribution in a dictionary is similar for a single NR block type as introduced in section \ref{sec:NRshape}. This distribution for each atom in the neighborhood is also similar within a single dictionary, but varies in magnitude based on the location due to the properties of the DCT. Fig. \ref{fig:nbd_corr} illustrates the correlation distribution of the atom at (3,3) position in the neighborhood for an exemplary dictionary of each block type. Diagonal symmetry (Types: 1,2,4) exhibits a near-perfect checkerboard pattern independent of the aspect ratio of the block. The dimension along which the polygon resembles the rectangle more (perpendicular to the parallel lines) exhibits higher correlation (Type: 3,5). The overall correlation in the dictionary is higher for types: 1,2,3 and therefore, the TX coefficients have more dissimilar properties to those of the DCT; thereby, designing an efficient entropy coding scheme is more crucial. 

\SubSection{Properties of the TX coefficients}
\label{sec:txprop}
 The TX coefficients have the following properties:
 \begin{enumerate}
  \item $l_0$ sparse or significantly high number of zeros;
  \item The highly correlated atoms produce TX coefficients that are unlikely to be non-zero together;
  \item The coefficients are highly correlated with their neighbors where the atoms are orthogonal;
  \item High concentration of energy in the top left corner.
 \end{enumerate} 
The first two properties are unique to this TX technique. In contrast, the last two properties are also observed in DCT coefficients and have already been utilized in the existing entropy coding scheme in AV1.


\Section{Proposed entropy coding}
\label{sec:Proposedmethod}
The current work aims to design an entropy coding scheme that effectively combines all the TX coefficient properties described in section \ref{sec:txprop}. The zig-zag scan and decomposition into 4 symbols, as described in section \ref{sec:av1entropycoding}, are maintained due to similar properties. The key novelty is in designing the contexts for NR coefficients. This section starts with designing a full context tree by extracting relevant information from already decoded symbols. This is followed by a scheme to simplify the context tree. However, these trees are designed for a single position. In the last section, a design to combine different positions for all NR shapes is described. This design is to be integrated into the AOM Video Model (AVM).\par
\SubSection{Full context tree (CT-f)}
\label{sec:full_context}
This section introduces a novel context design technique that combines the extracted conditions $\mathcal{C}_1, \mathcal{C}_2$ and $\mathcal{C}_3$ from the already decoded symbols in an immediate bottom-right neighborhood $\mathcal{N}_t$ of the current pixel. The neighborhood is defined by locations where the $l_1$ distance is $\leq n_\mathrm{nbd}$. A correlation is observed among neighboring TX coefficients, when the corresponding atoms are orthogonal or nearly orthogonal. Unlike DCT, certain non-zero neighboring coefficients reduce the likelihood of a non-zero coefficient at the current location. To parameterize these two properties, the $\mathcal{N}_t$ is divided into two neighborhood: a correlated neighborhood $\mathcal{N}_c$ (where the located atom has correlation $\geq \mathrm{th}_c$) and an orthogonal neighborhood $\mathcal{N}_o$ ($=\mathcal{N}_t \setminus\mathcal{N}_c$). An example of these neighborhoods is depicted in Fig. \ref{fig:example}-c. 
For simplicity, the set of coefficients and the size associated with a neighborhood $\mathcal{N}$ are defined as $|\mathcal{N}|$ and $\mathbf{x}_{\mathcal{N}}$, respectively. The conditions are:
\begin{itemize}
  \item $\mathcal{C}_1:$ $||\mathbf{x}_{\mathcal{N}_t}||_0= 0$, if $\mathcal{N}_t$ has only zeros
  \item $\mathcal{C}_2:$ $||\mathbf{x}_{\mathcal{N}_c}||_0$, number of non-zero coefficients in $\mathcal{N}_c$
  \item $\mathcal{C}_3:$ $||\mathbf{x}_{\mathcal{N}_o}||_1$, $l_1$ norm of the coefficients in $\mathcal{N}_o$
\end{itemize}
The outcomes of these conditions are concatenated into successive branches to model the context in a full tree as shown in Fig. \ref{fig:example}-a. The path to each leaf of the tree defines a unique context. If $\mathcal{C}_1$ is zero, the current coefficient is likely to be zero. Otherwise, other conditions are checked. $\mathcal{C}_2$ will have values from \{0,1,...,$|\mathcal{N}_c|$\}, which is location-dependant. When an entire $\mathcal{N}_c$ is occupied, the current coefficient is likely to be zero. In this case, this condition is sufficient for the context. On the other hand, when $\mathcal{C}_2$ is zero, the neighborhood behaves similarly to DCT. $\mathcal{C}_3$ is defined to have values \{0,1,...,12\}. If $\mathcal{C}_2$ is zero, it is not possible for $\mathcal{C}_3$ to be zero. The total number of contexts is equal to the number of leaves : \vxv{(||\mathbf{x}_{\mathcal{N}_c}||_0+1)}{13} when $|\mathcal{N}_c|<3$, otherwise \vxv{||\mathbf{x}_{\mathcal{N}_c}||_0}{13}$ + 1$.
\SubSection{Context tree merging (CT-m)}
\label{sec:merged_context}
A CT-f generally is highly complex. This section describes a method to merge the CT-f to reduce the complexity. The theoretical lower bound of the rate for a system with a set of contexts $\{\mathcal{S}_1\}$ and a set of symbols $\{\mathcal{S}_2\}$ is the conditional entropy $H(\mathcal{S}_2|\mathcal{S}_1) = -\sum_{i \in \mathcal{S}_2} \sum_{j \in \mathcal{S}_1} p(i,j)\log p(i|j)$. A merging of leaves at $\mathcal{C}_3$ is applied for each node at $\mathcal{C}_2$ to simplify the design. Starting from the first leaf, two neighboring leaves are successively merged if the loss of conditional entropy of the overall system remains below a threshold $\Delta$. If the neighbors can not be merged, the process moves to the next leaf and applies the same criteria. The lower bound of the number of contexts is $|\mathcal{N}_c|+1$.
\SubSection{Context simplification for integration (CT-s)}
\label{sec:sim_context}
The analysis of NR shapes in section \ref{sec:NRshape} motivated the focus on NR shapes with square and 1:2 bounding box and types 1-3. This leads to 9 unique shapes. A unique context map is designed for these 9 cases by combining the context map defined for DCT and the correlated position map essential to the context tree design. A merged tree is trained for the positions with the same number in the context map. The design and conditional probabilities will be integrated into the reference software to code TX coefficients for the NR region. Fig. \ref{fig:example}-e illustrates two such cases. Some slight modifications are applied to simplify the design, such as combining a few groups together and treating a few positions with a $|\mathcal{N}_c|$ differing by $\pm 1$ equivalently. Instead of designing a dynamic $\mathcal{N}_c$ for each case, the three most commonly used $|\mathcal{N}_c|=3$ and $|\mathcal{N}_c|=2$ as illustrated in Fig. \ref{fig:example}-d are stored, along with a lookup table to access them efficiently.
\Section{Experimental Setup and Results}
\label{sec:results}
\SubSection{Data:}
The NR TX method has been implemented in the AVM research-v2.0.0 \cite{AVMcode} software. 17 frames of Class A3, A4 and A5 videos from AOM Common Testing Conditions \cite{AVMcommon} are encoded. The TX coefficients during the trial encoding from frames 3, 7, 11 and 15 are used as the dataset. These frames are chosen due to the expected larger motions. The data for each case is randomly divided into 8:2 ratio for training and testing, respectively. 
      

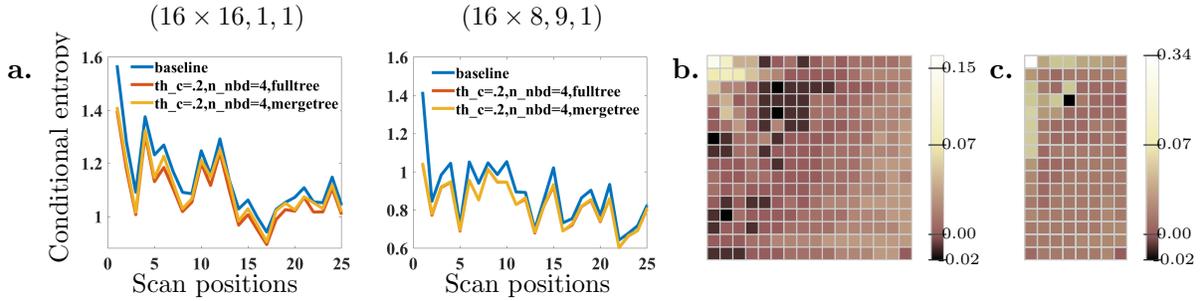
\begin{figure}[t]
   \setlength{\tabcolsep}{-2pt}
   \setlength\extrarowheight{-15pt}
 	\renewcommand{\arraystretch}{0}
  \centering
 	\begin{tabular}{p{.5cm}p{.4cm}p{4.2cm}p{4.2cm}p{.5cm}p{4cm}p{.5cm}p{4cm}}
   & & \raisebox{-.8cm}{\centerline{\footnotesize{($16\times16,1,1$)}}}&  \raisebox{-.8cm}{\centerline{\footnotesize{($16\times8,9,1$)}}}&&\\
    \hspace*{-.2cm}\raisebox{2.6cm}{\small{\textbf{a.}}}&\raisebox{1.6cm}{\rotatebox[origin=c]{90}{\footnotesize{Conditional entropy}}}&\includegraphics[width=4cm]{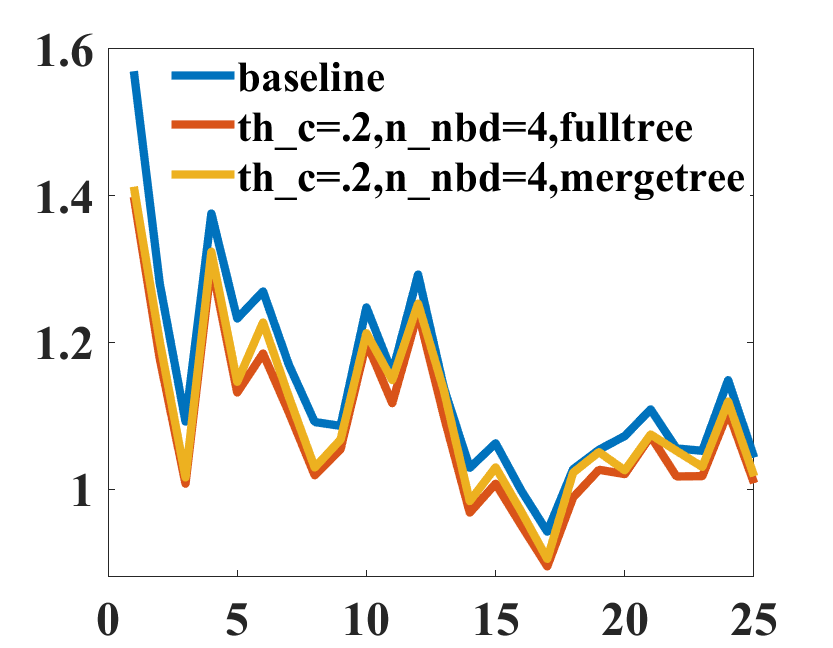}&\includegraphics[width=4cm]{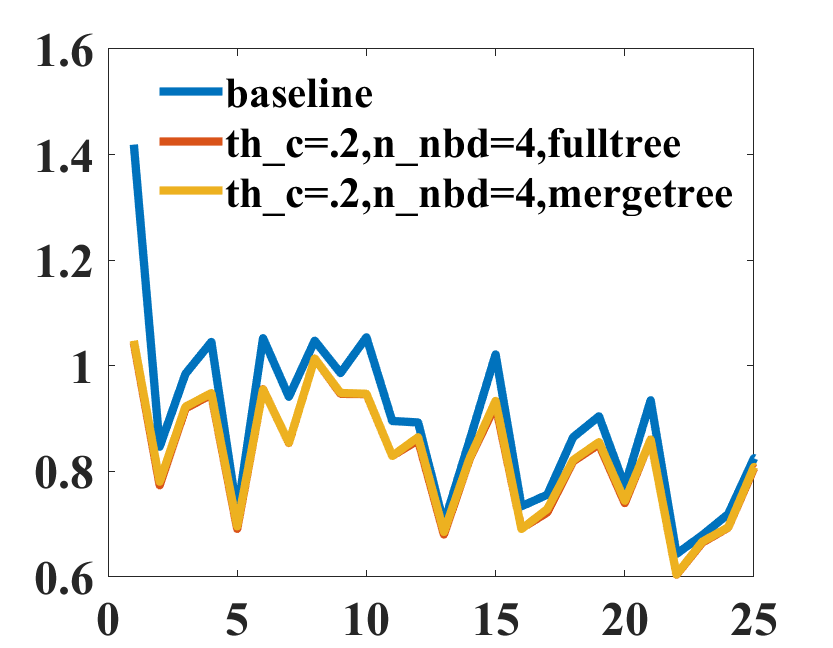} &\hspace*{-.1cm}\raisebox{2.6cm}{\small{\textbf{b.}}}&\begin{tikzpicture}[scale=0.17]
\newcommand\Square[1]{+(-#1,-#1) rectangle +(#1,#1)}

\foreach \y [count=\n] in {  
   {  0,     0,    57,    57 ,    0,     0,    14,     0,     0,    14,     0,    14,    14,    14,    14,    29},
   {100,     0,   100,    71 ,   43,    29,     0,     0,     0,     0,     0,    14,    29,    29,    29,    29},
   { 71,    29,    29,     0 ,    0,     0,     0,     0,     0,     0,     0,     0,     0,    14,    14,    29},
   { 29,    57,     0,    14 ,    0,     0,     0,     0,    14,    14,     0,    14,    14,    29,    29,    43},
   {  0,    71,    29,     0 ,    0,     0,     0,     0,    14,    14,    14,    14,    29,    29,    29,    29},
   {  0,    14,    43,     0 ,    0,    29,     0,     0,    14,     0,     0,    14,    14,    29,    29,    43},
   {  0,     0,    14,    14 ,   29,     0,     0,     0,     0,     0,    14,    14,    14,    43,    29,    43},
   {  0,     0,    14,    14 ,    0,     0,     0,     0,     0,    14,    14,    14,    29,    29,    29,    43},
   {  0,     0,     0,    14 ,    0,     0,     0,     0,     0,    14,    14,    14,    29,    43,    43,    43},
   { 14,     0,     0,    14 ,   14,     0,     0,     0,     0,    14,    14,    14,    29,    29,    43,    29},
   { 14,     0,    14,     0 ,    0,     0,     0,     0,    14,    14,    14,    29,    29,    29,    43,    43},
   {  0,     0,     0,     0 ,    0,     0,    14,     0,    14,    29,    29,    29,    43,    43,    43,    43},
   {  0,     0,    14,     0 ,    0,     0,     0,    14,    14,    14,    14,    29,    29,    29,    43,    43},
   {  0,     0,     0,     0 ,    0,     0,    14,    14,    14,    14,    29,    29,    29,    29,    29,    43},
   {  0,     0,     0,    14 ,   14,    29,    29,    29,    29,    43,    43,    43,    43,    43,    43,    43},
   {  0,     0,     0,     0 ,    0,     0,    14,    14,    14,    14,    29,    29,    29,    43,    43,     0},
   } {
      \foreach \x [count=\m] in \y {
        \draw[gray!40!,fill=yellow_heatmap!\x!pink_heatmap_m] (\m,-\n) \Square{14pt} ;
         
      }
    }


    \foreach \y [count=\n] in {  
   {  0,     0,     0,     0,    50,     0,     0,    50,     0,     0,     0,     0 ,    0,     0 ,    0 ,    0},
   {  0,     0,     0,     0,     0,     0,     0,     0,     0,     0,     0,     0 ,    0,     0 ,    0 ,    0},
   {  0,     0,     0,     0,    50,   100,    50,    50,    50,    50,     0,     0 ,    0,     0 ,    0 ,    0},
   {  0,     0,     0,     0,    50,    50,    50,    50,     0,     0,     0,     0 ,    0,     0 ,    0 ,    0},
   {  0,     0,     0,     0,    50,   100,    50,    50,     0,     0,     0,     0 ,    0,     0 ,    0 ,    0},
   {  0,     0,     0,     0,    50,     0,    50,    50,     0,     0,     0,     0 ,    0,     0 ,    0 ,    0},
   {100,    50,     0,     0,     0,    50,     0,     0,     0,     0,     0,     0 ,    0,     0 ,    0 ,    0},
   { 50,    50,     0,     0,    50,     0,     0,     0,     0,     0,     0,     0 ,    0,     0 ,    0 ,    0},
   {  0,     0,     0,     0,     0,    50,     0,     0,     0,     0,     0,     0 ,    0,     0 ,    0 ,    0},
   {  0,     0,     0,     0,     0,     0,     0,     0,     0,     0,     0,     0 ,    0,     0 ,    0 ,    0},
   {  0,     0,     0,     0,     0,     0,     0,     0,     0,     0,     0,     0 ,    0,     0 ,    0 ,    0},
   {  0,    50,     0,    50,    50,     0,     0,     0,     0,     0,     0,     0 ,    0,     0 ,    0 ,    0},
   { 50,   100,     0,     0,     0,     0,     0,     0,     0,     0,     0,     0 ,    0,     0 ,    0 ,    0},
   { 50,     0,     0,    50,     0,     0,     0,     0,     0,     0,     0,     0 ,    0,     0 ,    0 ,    0},
   {  0,     0,     0,     0,     0,     0,     0,     0,     0,     0,     0,     0 ,    0,     0 ,    0 ,    0},
   { 50,    50,    50,     0,     0,     0,     0,     0,     0,     0,     0,     0 ,    0,     0 ,    0 ,    0},
    } {
      \foreach \x [count=\m] in \y {
         \ifthenelse{\x>0}{\draw[gray!40!,fill=black!\x!pink_heatmap_m] (\m,-\n) \Square{14pt} ;}{}
         
      }
    }

  \foreach \y [count=\n] in {  
    {80,     25,     0,     0,     0,     0,     0,     0,     0,     0,     0,     0 ,    0 ,    0 ,    0,     0},
    { 0,     12,     0,     0,     0,     0,     0,     0,     0,     0,     0,     0 ,    0 ,    0 ,    0,     0},
    { 0,     0,     0,     0,     0,     0,     0,     0,     0,     0,     0,     0 ,    0 ,    0 ,    0,     0},
    { 0,     0,     0,     0,     0,     0,     0,     0,     0,     0,     0,     0 ,    0 ,    0 ,    0,     0},
    { 0,     0,     0,     0,     0,     0,     0,     0,     0,     0,     0,     0 ,    0 ,    0 ,    0,     0},
    { 0,     0,     0,     0,     0,     0,     0,     0,     0,     0,     0,     0 ,    0 ,    0 ,    0,     0},
    { 0,     0,     0,     0,     0,     0,     0,     0,     0,     0,     0,     0 ,    0 ,    0 ,    0,     0},
    { 0,     0,     0,     0,     0,     0,     0,     0,     0,     0,     0,     0 ,    0 ,    0 ,    0,     0},
    { 0,     0,     0,     0,     0,     0,     0,     0,     0,     0,     0,     0 ,    0 ,    0 ,    0,     0},
    { 0,     0,     0,     0,     0,     0,     0,     0,     0,     0,     0,     0 ,    0 ,    0 ,    0,     0},
    { 0,     0,     0,     0,     0,     0,     0,     0,     0,     0,     0,     0 ,    0 ,    0 ,    0,     0},
    { 0,     0,     0,     0,     0,     0,     0,     0,     0,     0,     0,     0 ,    0 ,    0 ,    0,     0},
    { 0,     0,     0,     0,     0,     0,     0,     0,     0,     0,     0,     0 ,    0 ,    0 ,    0,     0},
    { 0,     0,     0,     0,     0,     0,     0,     0,     0,     0,     0,     0 ,    0 ,    0 ,    0,     0},
    { 0,     0,     0,     0,     0,     0,     0,     0,     0,     0,     0,     0 ,    0 ,    0 ,    0,     0},
    { 0,     0,     0,     0,     0,     0,     0,     0,     0,     0,     0,     0 ,    0 ,    0 ,    0,     0},
     } {
      \foreach \x [count=\m] in \y {
         \ifthenelse{\x>0}{\draw[gray!40!,fill=white!\x!yellow_heatmap] (\m,-\n) \Square{14pt} ;}{}
         
      }
    }

\draw[white,fill, top color=pink_heatmap_m, bottom color=black] (18,-16.5) -- (19,-16.5) -- (19,-14.5) -- (18,-14.5) -- (18,-16.5);
\draw[white,fill, top color=yellow_heatmap, bottom color=pink_heatmap_m] (18,-14.5) -- (19,-14.5) -- (19,-7.5) -- (18,-7.5) -- (18,-14.5);
\draw[white,fill, top color=white, bottom color=yellow_heatmap] (18,-7.5) -- (19,-7.5) -- (19,-0.5) -- (18,-0.5) -- (18,-7.5);
\draw[gray!40] (18,-16.5) -- (19,-16.5) -- (19,-0.5) -- (18,-0.5) -- (18,-16.5);
\draw[] (17.8,-16.5) -- (19.2,-16.5);
\node at (20.2,-16.4) {\tiny{-0.02}};

\draw[] (17.8,-14.5) -- (19.2,-14.5);
\node at (20.2,-14.4) {\tiny{0.00}};

\draw[] (17.8,-7.5) -- (19.2,-7.5);
\node at (20.2,-7.4) {\tiny{0.07}};

\draw[] (17.8,-1.5) -- (19.2,-1.5);
\node at (20.2,-1.4) {\tiny{0.15}};


    \end{tikzpicture} & \hspace*{-.1cm}\raisebox{2.6cm}{\small{\textbf{c.}}}&\begin{tikzpicture}[scale=0.17]
\newcommand\Square[1]{+(-#1,-#1) rectangle +(#1,#1)}
 
\foreach \y [count=\n] in { 
   {  0,    71,     0 ,    0 ,   71 ,   71,    57 ,   43},
   { 57,    29,    14 ,   29 ,   29 ,   29,    29 ,   43},
   {  0,    29,    14 ,    0 ,   14 ,   14,    14 ,   14},
   {  0,    71,     0 ,    0 ,   29 ,   14,    14 ,    0},
   {  0,    57,    57 ,   14 ,   14 ,   14,    14 ,   14},
   {  0,    29,    14 ,   14 ,   14 ,   29,    29 ,   14},
   {  0,    29,    14 ,   14 ,   14 ,   14,    29 ,   14},
   {  0,    14,    29 ,   29 ,   29 ,   29,    29 ,   14},
   { 57,    14,    14 ,   14 ,   29 ,   29,    29 ,   29},
   { 43,    14,    14 ,   29 ,   29 ,   29,    29 ,   29},
   { 43,    14,    29 ,   29 ,   29 ,   29,    29 ,   29},
   { 29,    14,    29 ,   29 ,   29 ,   29,    43 ,   29},
   { 43,    29,    29 ,   29 ,   29 ,   29,    43 ,   29},
   { 29,    29,    29 ,   29 ,   29 ,   29,    43 ,   43},
   { 29,    29,    29 ,   29 ,   43 ,   43,    43 ,   43},
   { 14,    29,    29 ,   29 ,   29 ,   29,    43 ,    0},
    } {
     \foreach \x [count=\m] in \y {
        \draw[gray!40!,fill=yellow_heatmap_m!\x!pink_heatmap_m] (\m,-\n) \Square{14pt};
     }
   }

  \foreach \y [count=\n] in { 
     {0,     0 ,    0,     0,     0 ,    0 ,    0 ,    0},
     {0,     0 ,    0,     0,     0 ,    0 ,    0 ,    0},
     {0,     0 ,    0,     0,     0 ,    0 ,    0 ,    0},
     {0,     0 ,    0,    100,     0 ,    0 ,    0 ,    0},
     {0,     0 ,    0,     0,     0 ,    0 ,    0 ,    0},
     {0,     0 ,    0,     0,     0 ,    0 ,    0 ,    0},
     {0,     0 ,    0,     0,     0 ,    0 ,    0 ,    0},
     {0,     0 ,    0,     0,     0 ,    0 ,    0 ,    0},
     {0,     0 ,    0,     0,     0 ,    0 ,    0 ,    0},
     {0,     0 ,    0,     0,     0 ,    0 ,    0 ,    0},
     {0,     0 ,    0,     0,     0 ,    0 ,    0 ,    0},
     {0,     0 ,    0,     0,     0 ,    0 ,    0 ,    0},
     {0,     0 ,    0,     0,     0 ,    0 ,    0 ,    0},
     {0,     0 ,    0,     0,     0 ,    0 ,    0 ,    0},
     {0,     0 ,    0,     0,     0 ,    0 ,    0 ,    0},
     {0,     0 ,    0,     0,     0 ,    0 ,    0 ,    0},
     } {
      \foreach \x [count=\m] in \y {
         \ifthenelse{\x>0}{\draw[gray!40!,fill=black] (\m,-\n) \Square{14pt} ;}{}
         
      }
    }

\foreach \y [count=\n] in { 
   {108,     0 ,    4 ,    4 ,    0 ,    0 ,    0 ,    0},
   {  0,     0 ,    0 ,    0 ,    0 ,    0 ,    0 ,    0},
   {  4,     0 ,    0 ,    8 ,    0 ,    0 ,    0 ,    0},
   {  8,     0 ,    8 ,    0 ,    0 ,    0 ,    0 ,    0},
   { 23,     0 ,    0 ,    0 ,    0 ,    0 ,    0 ,    0},
   { 15,     0 ,    0 ,    0 ,    0 ,    0 ,    0 ,    0},
   { 12,     0 ,    0 ,    0 ,    0 ,    0 ,    0 ,    0},
   { 12,     0 ,    0 ,    0 ,    0 ,    0 ,    0 ,    0},
   {  0,     0 ,    0 ,    0 ,    0 ,    0 ,    0 ,    0},
   {  0,     0 ,    0 ,    0 ,    0 ,    0 ,    0 ,    0},
   {  0,     0 ,    0 ,    0 ,    0 ,    0 ,    0 ,    0},
   {  0,     0 ,    0 ,    0 ,    0 ,    0 ,    0 ,    0},
   {  0,     0 ,    0 ,    0 ,    0 ,    0 ,    0 ,    0},
   {  0,     0 ,    0 ,    0 ,    0 ,    0 ,    0 ,    0},
   {  0,     0 ,    0 ,    0 ,    0 ,    0 ,    0 ,    0},
   {  0,     0 ,    0 ,    0 ,    0 ,    0 ,    0 ,    0},
    } {
     \foreach \x [count=\m] in \y {
        \ifthenelse{\x>0}{\draw[gray!40!,fill=white!\x!yellow_heatmap_m] (\m,-\n) \Square{14pt} ;}{}
        \draw[line width=0.08mm,gray!40!] (\m,-\n) \Square{14pt} ;
     }
   }

\draw[white,fill, top color=pink_heatmap_m, bottom color=black] (10,-16.5) -- (11,-16.5) -- (11,-14.5) -- (10,-14.5) -- (10,-16.5);
\draw[white,fill, top color=yellow_heatmap, bottom color=pink_heatmap_m] (10,-14.5) -- (11,-14.5) -- (11,-7.5) -- (10,-7.5) -- (10,-14.5);
\draw[white,fill, top color=white, bottom color=yellow_heatmap] (10,-7.5) -- (11,-7.5) -- (11,-0.5) -- (10,-0.5) -- (10,-7.5);
\draw[gray!40] (10,-16.5) -- (11,-16.5) -- (11,-0.5) -- (10,-0.5) -- (10,-16.5);
\draw[] (9.8,-16.5) -- (11.2,-16.5);
\node at (12.2,-16.4) {\tiny{-0.02}};

\draw[] (9.8,-14.5) -- (11.2,-14.5);
\node at (12.2,-14.4) {\tiny{0.00}};

\draw[] (9.8,-7.5) -- (11.2,-7.5);
\node at (12.2,-7.4) {\tiny{0.07}};

\draw[] (9.8,-.5) -- (11.2,-.5);
\node at (12.2,-.4) {\tiny{0.34}};

\end{tikzpicture}\\
    &\hspace*{2cm}\centerline{\footnotesize{Scan positions}}&  \hspace*{4cm}\centerline{\footnotesize{Scan positions}} &&&&\\
      
 	\end{tabular}	
  \caption{a. Full and merged tree of first 25 positions, b. $\Delta H_{\mathrm{ts}}$ heatmap ($16\times 16,1,1$), c. $\Delta H_{\mathrm{ts}}$ heatmap ($16\times 8,9,1$)}
   	\label{fig:full_merge_res}
\end{figure}

\SubSection{Experiments:}
\begin{table}[b]
  \centering
  \caption{number of contexts and entropy savings} \label{tab:ctx_entropy}
  \vspace{-0.2cm}
  \setlength{\tabcolsep}{4pt} 
  \hspace*{.4cm}
\begin{tabular}{lccccccc}
  \toprule
  $\mathrm{B_d}$ & Type &$\#\mathrm{ctx}_{\mathrm{aom}}$ &$\#\mathrm{ctx}$ & $\Delta H$ & $\Delta H_\mathrm{tl}$&\#$\mathrm{NP}$& \#$\mathrm{NP_{tl}}$\\
  \midrule
  $4,8$&2&4&6&.65&.51&1&1\\
  $8,8$&3&4&8&1.38&.93&2&0\\
  $8,16$&1&4&4&3.68&2.57&1&1\\
  $8,16$&2&4&4&1.41&.68&10&10\\
  $8,16$&3&4&8&3.13&2.15&5&5\\
  $16,16$&3&4&9&2.84&.94&36&33\\
  \bottomrule
\end{tabular}
\end{table}
The entropy coding scheme from AV1 is implemented outside the software as a baseline for comparison. The training data set is used to calculate the conditional probabilities \{$p_{\mathrm{tr}}(i|j)$\} for both the proposed schemes and the baseline. Additionally, the contexts \{$\mathcal{S}_1$\} are established for the merged tree. 
The \{$p_{\mathrm{tr}}(i|j)$\} and \{$\mathcal{S}_1$\} are used to calculate the conditional entropy $H_{\mathrm{ts}}=-\sum_{i \in \mathcal{S}_2} \sum_{j \in \mathcal{S}_1} p_{\mathrm{ts}}(i,j)\log p_{\mathrm{tr}}(i|j)$ for \{$\mathcal{S}_2$\} of BR symbols in the test data set for both schemes for evaluation, where $p_{\mathrm{ts}}(i,j)$ is determined in the test dataset.\par
A CT-f and CT-m are designed for each position, covering all resultant NR shapes with dimensions in \{8,16\} using the training data. The baseline's \{$p_{\mathrm{tr}}(i|j)$\} are calculated for each position separately for a fair comparison.  From the tested parameters $n_\mathrm{nbd} =\{2, 3, 4\}$ and $\mathrm{th}_c = \{0.09, 0.15, 0.2, 0.25\}$, results indicate that $n_\mathrm{nbd}=4$ and $\mathrm{th}_c =0.2$ performs best. Fig. \ref{fig:full_merge_res}-a presents 2 examples of position-wise $H_{\mathrm{ts}}$ in the zig-zag scanned vector for Type 3 ($16\times 16,1,1$) and Type 1 ($16\times 8,9,1$). The total block count in the dataset is $\sim\!70$K and $\sim\!\!\!140$K respectively. The proposed methods improve performance across all examined positions. The merged tree for ($16\times 16,1,1$) typically has contexts between 5 and 9 with minimal loss compared to the full tree, while the tree for ($16\times 8,9,1$) has either 6 or 7 contexts with negligible loss.\par
The results for the same blocks and datasets using the $\text{CT-s}$ are shown as an example in Fig. \ref{fig:full_merge_res}-b,c. The $\{p_{\mathrm{tr}}(i|j)\}$ of the baseline is calculated for the positions with identical contexts together using the AV1 context map. The results for Types 1-3 are discussed in this section. Types 4, 5 are also tested, but they provided minimal gains, which is expected. The figure illustrates the difference in $H_{\mathrm{ts}}$ between the baseline and the simplified scheme ($\Delta H_{\mathrm{ts}}$). Positive values imply an improved performance. For ($16\times 8,9,1$), Fig. \ref{fig:full_merge_res}-c, gains are observed for all positions except one, while ($16\times 16,1,1$), Fig. \ref{fig:full_merge_res}-b shows improved performance mainly in the top-left corner and bottom-right diagonal half. A slight loss is noted in some mid-frequency positions within the area indicated between the lines in Fig. \ref{fig:full_merge_res}-b. Table \ref{tab:ctx_entropy} presents the overall results for the Type 1-3 NR shapes. It presents the parameters: \#$\text{ctx}_{\text{aom}}$: number of contexts in aom, \#$\text{ctx}$: number of contexts in CT-s, $\Delta H:$ sum of $\Delta H_{\mathrm{ts}}$ for all positions, $\Delta H_{\text{tl}}$: the same in the top-left diagonal region, \#NP: number of locations where a loss is observed, \#NP$_{\text{tl}}$: the same in the top-left diagonal region. There is a consistent gain observed for all the cases. The \vxv{16}{16} Type 3 is the case where more losses are observed. \par
\Section{Conclusion}
\label{sec:concl}
This work presented a novel entropy coding design for NR transformation using partitioned DCT and OMP. The design provides performance improvement for block types where the transform coefficients behave very differently than those of DCT coefficients. The design is to be implemented in the reference software for these block types as a part of future work. Future work also includes testing possible design improvements such as considering the $l_1$ norm in  the correlated neighborhood and different scan patterns for the block types closer to a rectangle.



\Section{References}
\bibliographystyle{IEEEtran}
\bibliography{reference}

\end{document}